\documentclass{revtex4}
\usepackage{amssymb}
\usepackage{amsfonts}
\usepackage{amsmath}

\input{tcilatex}

\begin{document}

\preprint{}
\title{Stable Magnetic Universes Revisited}
\author{T. Tahamtan}
\email{tayabeh.tahamtan@emu.edu.tr}
\author{M. Halilsoy }
\email{mustafa.halilsoy@emu.edu.tr }
\affiliation{Department of Physics, Eastern Mediterranean University, G. Magusa, North
Cyprus, Mersin 10 - Turkey.}

\begin{abstract}
A regular class of static, cylindrically symmetric pure magnetic field
metrics is rederived in a different metric ansatz in all dimensions. Radial,
time dependent perturbations show that for dimensions $d>3$ such spacetimes
are stable at both near $r\approx 0$ and large radius $r\rightarrow \infty $%
. In a different gauge these stability analysis and similar results were
known beforehand. For $d=3$, however, simultaneous stability requirement at
both, near and far radial distances can not be reconciled for time -
dependent perturbations. Restricted, numerical geodesics for neutral
particles reveal a confinement around the center in the polar plane.
Charged, time-like geodesics for $d=4$ on the other hand are shown
numerically to run toward infinity.
\end{abstract}

\pacs{PACS number}
\maketitle

\section{Introduction}

In $4-$dimensional spacetimes the electric - magnetic duality symmetry of
the Maxwell equations is an important property which can not be satisfied in
other dimensions unless different form fields other than $2-$forms are
introduced. For this reason a dyonic solution admits a meaningful
interpretation only in $d=4$. The Reissner - Nordestrom (RN) solution
constitutes in this regard the best example which has both electric and
magnetic solutions in a symmetric manner. In other dimensions ( $d\neq 4$ )
similar duality properties can, in principle, be defined as well but
physical interpretation corresponding to electric and magnetic fields turn
out to be rather abstract. For such reasons, in order\ to avoid
complications due to the absence of a tangible duality, pure electric or
pure magnetic solutions seemed to attract considerable attention. This
amounts to only half of the Maxwell equations, the other half being
trivially satisfied. From this token we wish to resort here to the pure
magnetic solutions which yield a completely solvable class without much
effort. From the physical side, occurrence of pure and very strong magnetic
fields associated with astronomical objects such as magnetars motivate us to
search for such solutions in general relativity.

It was Melvin, who first studied such cylindrically symmetric parallel
magnetic lines of force remaining in equilibrium under their mutual
gravitational attraction in $d=4$\cite{1}. Later on, generalized version of
the Melvin's magnetic universe was also considered\cite{2}. The Melvin
universe is invariant under rotation and translation along the $z_{i}$ axis
orthogonal to the polar plane $\left( r,\varphi \right) $. Thorne
popularized the Melvin universe further by showing its absolute stability
against small radial perturbations\cite{3}. Due to this stability property
it can be presumed that astrophysical objects emitting strong beams of
magnetic fields may everlast in an accelerating universe. Additionally, in $%
d=3$ \cite{4} and $5$ dimensional \cite{5} cases also pure magnetic field
solutions were found and their energy content investigated \cite{6}. Pure
magnetic solutions in higher dimensions are also known to exist in string,
Lovelock, Yang-Mills, Born-Infeld and other theories \cite{7}.

In this paper, we present in a particular cylindrically symmetric metric
ansatz, a class of non-singular, source-free, static, pure magnetic
solutions to Einstein-Maxwell (EM) equations in all dimensions. In a
different metric ansatz these solutions were known previously \cite{4,8}.
Our principal aim is to investigate the stability of such magnetic universes
against time dependent small radial perturbations and explore the possible
role of dimensionality of spacetime in such matters. It has been known for a
long time that for $d>3$ these kind of magnetic solutions are all stable 
\cite{8} . We verify these results once more in a different metric (i.e. non
canonical) ansatz with supplement of the $d=3$ case.. We show that for $d>3$
the metrics are stable against small perturbations at both near axis $z_{i}$
and at far distance away from $z_{i}$. We observe also that when $d=3$ these
two regions behave differently. Namely, the metric can be made stable at $%
r\approx 0$ or, at $r\rightarrow \infty $, but not simultaneously, which we
phrase as 'weakly' stable. The solutions justify once more the impossibility
of cylindrical magnetic field lines implosion and therefore formation of
such black holes. We investigate the time-like ( $d\geq 4$,with fixed polar
angle) and null ( $d=3,4$ ) geodesics for neutral particles numerically.
Only for $d=4$ and $5$ we were able to obtain exact integrals, albeit in
non-invertible forms, of the geodesics equation. In each case a confinement
of geodesics is observed to take place near the central region. Due to its
physical importance we consider also the time-like geodesics of a charged
particle. It turns out that such geodesics can not be confined and in their
proper time they diverge to infinity.

Organization of the paper is as follows. In Section II\ we present our
metric, field equations and solve them in $d-$dimensions. Perturbation
analysis of our system follows in Section III. Geodesics motion is studied
in Section IV. Our results are summarized in Conclusion which appears in
Section V.

\section{Metric and solutions of field equations in $d-$dimensions}

Our $d-$dimensional static, cylindrically symmetric line element ansatz is
given by

\begin{equation}
ds^{2}=f\left( r\right) \left( dt^{2}-dr^{2}-\underset{i=1}{\overset{d-3}{%
\tsum }}dz_{i}^{2}\right) -\frac{r^{2}b_{0}^{2}}{f\left( r\right) ^{k}}%
d\varphi ^{2},\text{ \ \ \ \ \ \ \ \ \ \ \ \ \ \ \ \ \ \ \ \ \ \ \ \ \ \ \ }%
d\geq 3
\end{equation}%
in which $f\left( r\right) $ is a function of $r$ to be found and $b_{0}$
and $k$ are constant parameters. Also the pure magnetic $2-$form field is
chosen to be 
\begin{equation}
\mathbf{F=}F_{r\varphi }dr\wedge d\varphi ,
\end{equation}%
where $F_{r\varphi }$ is the only non-zero component of the electromagnetic
field. The energy momentum tensor is defined by

\begin{equation}
4\pi T_{i}^{j}=-F_{ik}F^{jk}+\frac{1}{4}\delta _{i}^{j}F_{mn}F^{mn}
\end{equation}%
\bigskip which admits the non-zero components%
\begin{equation}
T_{i}^{j}=\text{diag}\left[ T_{0}^{0}=T_{2}^{2}=T_{3}^{3}=\cdots
=-T_{1}^{1}=-T_{d-1}^{d-1}\right] =\frac{1}{8\pi }F_{r\varphi }F^{r\varphi }.
\end{equation}%
We note that our choice of indices $\left\{ 0,1,2,\ldots ,(d-1)\right\} $
denote $\left\{ t,r,z_{1},\cdots ,z_{d-3},\varphi \right\} $ and the energy
conditions satisfied by this energy-momentum tensor are discussed in the
Appendix A. From the Einstein equations, $T_{i}^{j}=G_{i}^{j}$ Eq. (4)
implies that 
\begin{equation}
G_{0}^{0}=G_{2}^{2}=G_{3}^{3}=\cdots =-G_{1}^{1}=-G_{d-1}^{d-1}.
\end{equation}%
\bigskip From these relations we can write

\begin{eqnarray}
G_{1}^{1} &=&G_{d-1}^{d-1} \\
G_{1}^{1} &=&-G_{0}^{0} \\
G_{0}^{0} &=&-G_{d-1}^{d-1}
\end{eqnarray}%
where

\begin{eqnarray}
G_{0}^{0} &=&\frac{1}{4f^{3}r}\left[ 2ff^{\text{ }\prime \prime }r\left(
k-\left( d-3\right) \right) +rf^{%
{\acute{}}%
\text{ }2}\left( -k^{2}+\left( d-6\right) k-\left( d-8\right) \frac{\left(
d-3\right) }{2}\right) +2ff%
{\acute{}}%
\left( 2k-\left( d-4\right) \right) \right] \\
G_{1}^{1} &=&\frac{1}{4f^{3}r}\left[ rf^{%
{\acute{}}%
\text{ }2}\left( \left( d-2\right) k-\left( d-2\right) \frac{\left(
d-3\right) }{2}\right) -2ff%
{\acute{}}%
\left( d-2\right) \right] \\
G_{d-1}^{d-1} &=&\frac{1}{4f^{3}r}\left[ -2ff^{\text{ }\prime \prime
}r\left( d-2\right) -rf\text{ }^{%
{\acute{}}%
\text{ }2}\left( \left( d-2\right) \frac{\left( d-7\right) }{2}\right) %
\right] .
\end{eqnarray}%
From equation (6) we obtain the differential equation 
\begin{equation}
2rf^{\text{ }\prime \prime }f+(k-2)rf%
{\acute{}}%
\text{ }^{2}-2ff%
{\acute{}}%
=0.
\end{equation}%
whose solution for $d\geq 4$ is%
\begin{equation}
f(r)=(kr^{2}+C_{1})^{\frac{2}{k}}.
\end{equation}%
By putting this result into Eq. (7) for finding $k$ we find out that for $%
d\geq 4$, $k$ is $d-3.$ On the other hand, for $d=3$ the solution turns out
to be 
\begin{equation}
f(r)=C_{2}e^{cr^{2}}
\end{equation}%
for the integration constants $C_{2}$ and $c$. For convenience we make the
choices $C_{1}=d-3$ and $C_{2}=1$, so that the solution can be expressed by

\begin{equation}
f(r)=\left\{ 
\begin{array}{cc}
(d-3)^{\frac{2}{d-3}}(r^{2}+1)^{\frac{2}{d-3}} & \text{ \ \ \ \ \ \ \ \ \ \
\ \ \ \ \ \ \ \ \ \ \ \ \ \ \ \ \ \ \ \ \ \ \ \ \ \ \ \ \ \ \ \ \ \ \ \ \ }%
d\geq 4 \\ 
e^{cr^{2}} & \text{ \ \ \ \ \ \ \ \ \ \ \ \ \ \ \ \ \ \ \ \ \ \ \ \ \ \ \ \
\ \ \ \ \ \ \ \ \ \ \ \ \ \ \ \ \ \ \ }d=3%
\end{array}%
\right. .
\end{equation}%
Accordingly, our line element takes the form

\begin{equation}
ds^{2}=\left\{ 
\begin{array}{cc}
(d-3)^{\frac{2}{d-3}}(r^{2}+1)^{\frac{2}{d-3}}\left( dt^{2}-dr^{2}-\underset{%
i=1}{\overset{d-3}{\tsum }}dz_{i}^{2}\right) -\frac{r^{2}b_{0}^{2}}{%
(d-3)^{2}(r^{2}+1)^{2}}d\varphi ^{2}, & \text{\ \ \ \ \ \ \ \ }d\geq 4 \\ 
e^{cr^{2}}\left( dt^{2}-dr^{2}\right) -r^{2}b_{0}^{2}d\varphi ^{2}, & \text{
\ \ \ \ \ \ \ }d=3%
\end{array}%
\right. .
\end{equation}%
We note that these solutions are not new, for they coincide with those of 
\cite{8} (for $d\geq 4$) and \cite{4} for ($d=3$), respectively. It can
easily be seen that for $r\rightarrow 0$ it reduces to the following form

\begin{equation}
ds^{2}\approx \left\{ 
\begin{array}{cc}
(d-3)^{\frac{2}{d-3}}(dt^{2}-dr^{2}-\underset{i=1}{\overset{d-3}{\tsum }}%
dz_{i}^{2})-\frac{r^{2}b_{0}^{2}}{(d-3)^{2}}d\varphi ^{2}, & \text{ \ \ \ \
\ \ \ \ \ \ \ \ \ \ \ \ \ \ \ \ \ \ \ \ \ \ \ \ \ \ \ \ \ \ \ \ }d\geq 4 \\ 
(dt^{2}-dr^{2})-r^{2}b_{0}^{2}d\varphi ^{2}, & \text{ \ \ \ \ \ \ \ \ \ \ \
\ \ \ \ \ \ \ \ \ \ \ \ \ \ \ \ \ \ \ \ \ \ \ \ \ }d=3%
\end{array}%
\right.
\end{equation}%
This represents a conical geometry signalling the existence of a cosmic
string near $r=0$. By choosing $\frac{b_{0}^{2}}{(d-3)^{2}}=1$ for $d\geq 4$
and $b_{0}=1$ for $d=3$ , we have Minkowskian metrics as one approaches the
axes $z_{i}$. The solution (16) is a singularity free magnetic universe in $%
d-$ dimensions in analogy with the Melvin space time.

From Maxwell's Eq. it fallows that

\bigskip 
\begin{eqnarray}
F^{r\varphi } &=&\left\{ 
\begin{array}{cc}
\frac{B_{0}}{(d-3)^{\frac{2}{d-3}}rb_{0}(1+r^{2})^{\frac{2}{d-3}}}, & \text{
\ \ \ \ \ \ \ \ \ \ \ \ \ \ \ \ \ \ \ \ \ \ \ \ \ \ \ \ \ \ \ \ \ \ \ \ }%
d\geq 4 \\ 
\frac{B_{0}}{rb_{0}}e^{-cr^{2}}, & \text{ \ \ \ \ \ \ \ \ \ \ \ \ \ \ \ \ \
\ \ \ \ \ \ \ \ \ \ \ \ \ \ \ \ \ \ \ }d=3%
\end{array}%
\right. \\
&&\text{(}B_{0}=\text{ an integration constant)}  \notag
\end{eqnarray}%
which implies that the magnetic field behaviors as a function of $r$ are

\begin{equation*}
F_{r\varphi }\sim \left\{ 
\begin{array}{c}
\frac{r}{(1+r^{2})^{2}}\text{\ \ \ \ \ \ \ \ \ \ \ \ \ \ \ \ \ \ \ \ \ \ \ \
\ \ \ \ \ \ \ \ \ \ }d\geq 4 \\ 
r\text{ \ \ \ \ \ \ \ \ \ \ \ \ \ \ \ \ \ \ \ \ \ \ \ \ \ \ \ \ \ \ \ \ \ \
\ \ \ \ \ \ \ \ }d=3%
\end{array}%
\right. .
\end{equation*}%
The marked distinction between $d=3$ and $d\geq 4$ cases can already be seen
from these behaviors. Accordingly the energy density reads

\begin{equation}
T_{0}^{0}=\left\{ 
\begin{array}{cc}
\frac{B_{0}^{2}}{8\pi (d-3)^{\frac{2d-4}{d-3}}(1+r^{2})^{\frac{2d-4}{d-3}}}
& \text{ \ \ \ \ \ \ \ \ \ \ \ \ \ \ \ \ \ \ \ \ \ \ \ \ \ \ \ \ \ \ \ \ }%
d\geq 4 \\ 
\frac{B_{0}^{2}}{8\pi }e^{-cr^{2}} & \text{ \ \ \ \ \ \ \ \ \ \ \ \ \ \ \ \
\ \ \ \ \ \ \ \ \ \ \ \ \ \ \ \ }d=3%
\end{array}%
\right.
\end{equation}%
while the Ricci scalar for the metric (16) is

\begin{equation}
R=\left\{ 
\begin{array}{cr}
\frac{-4(d-4)(d-3)^{^{\frac{1-d}{d-3}}}}{(1+r^{2})^{\frac{2(d-2)}{d-3}}} & 
\text{\ \ \ \ }d\geq 4 \\ 
2ce^{-cr^{2}} & \text{ \ \ \ \ \ \ }d=3%
\end{array}%
\right. .
\end{equation}%
Similarly, the Kretchmann scalar has the behavior

\begin{equation}
K\sim \left\{ 
\begin{array}{cc}
\frac{1}{(d-3)^{\frac{4(d-2)}{d-3}}(1+r^{2})^{\frac{4(d-2)}{d-3}}} & \text{
\ \ \ \ \ \ \ \ \ \ \ \ \ \ \ \ \ \ \ \ \ \ \ \ \ \ \ \ \ \ \ \ \ \ \ \ }%
d\geq 4 \\ 
12c^{2}e^{-2cr^{2}} & \text{ \ \ \ \ \ \ \ \ \ \ \ \ \ \ \ \ \ \ \ \ \ \ \ \
\ \ \ \ \ \ \ \ \ \ \ }d=3%
\end{array}%
\right. .
\end{equation}%
It is observed that regularity at $r\rightarrow \infty $ dictates us to make
the choice $c>0$ for the integration constant.

\section{\textbf{Perturbation Analysis}}

\bigskip In this section we perturb the metric and magnetic potential. Since
the case $d=3$ forms a special case we consider it separately. Similar
analysis was carried out by Gibbons and Wiltshire \cite{8} where they used
the canonical metric ansatz. We shall show below that their results can also
be obtained in a different metric ansatz.

\subsubsection{The case for $d=3$}

Our line element is

\begin{equation}
ds^{2}=f(r,t)\left( dt^{2}-dr^{2}\right) -\frac{r^{2}b_{0}^{2}}{g(r,t)}%
d\varphi ^{2}
\end{equation}%
where 
\begin{eqnarray}
f(r,t) &=&f_{0}(r)+\epsilon u(r,t) \\
g(r,t) &=&g_{0}(r)+\epsilon w(r,t)  \notag
\end{eqnarray}%
and the magnetic potential is expressed by

\begin{equation}
A_{\varphi }(r,t)=A_{\varphi }(r)+\epsilon a(r,t).
\end{equation}%
The unperturbed functions are

\begin{eqnarray}
f_{0}(r) &=&e^{cr^{2}}\text{ \ \ \ } \\
g_{0}(r) &=&1 \\
A_{\varphi }(r) &=&\frac{B_{0}b_{0}r^{2}}{2}
\end{eqnarray}%
where $u\left( r,t\right) ,w\left( r,t\right) $ and $a\left( r,t\right) $
are the perturbed functions. Since $\epsilon $ is a small parameter we
assume that $\epsilon ^{2}\approx 0$ in our analysis. We use Einstein's
equations to find the perturbed functions. The differential equations
satisfied by the perturbed functions are%
\begin{equation}
\left( cr^{2}\frac{\partial w(r,t)}{\partial r}-r\frac{\partial ^{2}w(r,t)}{%
\partial t^{2}}\right) e^{cr^{2}}+b_{0}\left[ r\frac{\partial ^{2}u(r,t)}{%
\partial r^{2}}-r\frac{\partial ^{2}u(r,t)}{\partial t^{2}}-4r^{2}c\frac{%
\partial u(r,t)}{\partial r}+4r^{3}c^{2}u(r,t)-\frac{\partial u(r,t)}{%
\partial r}\right] =0
\end{equation}

\begin{equation}
r\frac{\partial ^{2}w(r,t)}{\partial r^{2}}+2\frac{\partial w(r,t)}{\partial
r}-r\frac{\partial ^{2}w(r,t)}{\partial t^{2}}=0
\end{equation}%
\begin{equation}
-\frac{\partial ^{2}a(r,t)}{\partial t^{2}}+\frac{\partial ^{2}a(r,t)}{%
\partial r^{2}}-\frac{1}{r}\frac{\partial a(r,t)}{\partial r}+\frac{%
B_{0}b_{0}r}{2}\frac{\partial w(r,t)}{\partial r}=0.
\end{equation}%
\bigskip This system of differential equations admits the solutions%
\begin{eqnarray}
u(r,t) &=&e^{-\alpha t+cr^{2}}\left( \frac{4rB_{0}}{b_{0}}\left[
B_{1}I_{1}(\alpha r)+B_{2}K_{1}(\alpha r)\right] \right. + \\
&&cre^{(-\alpha r-\alpha )}\left\{ -(\sinh \alpha r+\cosh \alpha
r)(E_{1}+E_{2})e^{2\alpha r}+e^{2\alpha }(\sinh \alpha r-\cosh \alpha
r)(-E_{1}+E_{2})\right\} +  \notag \\
&&\left. \left\{ \left( crE_{2}-\alpha E_{1}\right) \cosh \alpha r+\left(
crE_{1}-\alpha E_{2}\right) \sinh \alpha r\right\} \right)  \notag
\end{eqnarray}

\begin{equation}
w(r,t)=e^{-\alpha t}(E_{1}\frac{\sinh \alpha r}{r}+E_{2}\frac{\cosh \alpha r%
}{r})
\end{equation}

\begin{eqnarray}
a(r,t) &=&re^{-\alpha t}\left( B_{1}I_{1}(\alpha r)+B_{2}K_{1}(\alpha
r)-\right. \\
&&\left. \frac{B_{0}b_{0}}{4}e^{-\alpha r-\alpha }\left\{ (\sinh \alpha
r+\cosh \alpha r)(E_{1}+E_{2})e^{2\alpha r}+e^{2\alpha }(\sinh \alpha
r-\cosh \alpha r)(-E_{1}+E_{2})\right\} \right) ,  \notag
\end{eqnarray}%
where $B_{1},B_{2},E_{1},E_{2}$ and $\alpha $ are all integration constants
while $I_{1}(\alpha r)$ and $K_{1}(\alpha r)$ are the modified Bessel
functions of order one. At all times $t$, to a first order in $\epsilon $ ,
the locally flat nature of the metric near $r=0$ will not be altered \cite{1}%
. This implies when $r\rightarrow 0$

\begin{eqnarray}
u(r,t) &=&e^{-\alpha t+cr^{2}}\left( \frac{4rB_{0}}{b_{0}}B_{2}K_{1}\left(
\alpha r\right) -crE_{1}e^{-\alpha r-\alpha }\left\{ (\sinh \alpha r+\cosh
\alpha r)e^{2r\alpha }+e^{2\alpha }(\sinh \alpha r-\cosh \alpha r)\right\}
\right. \\
&&\left. +E_{1}\left\{ cr\sinh \alpha r-\alpha \cosh \alpha r\right\}
\right) ,  \notag
\end{eqnarray}%
\begin{equation}
w(r,t)=e^{-\alpha t}E_{1}\frac{\sinh \alpha r}{r},
\end{equation}

\begin{equation}
a(r,t)=re^{-\alpha t}\left( B_{2}K_{1}(\alpha r)-\frac{b_{0}B_{0}}{4}%
E_{1}e^{-\alpha r-\alpha }\left\{ (\sinh \alpha +\cosh \alpha )e^{2\alpha
r}-e^{2\alpha }(\sinh \alpha -\cosh \alpha )\right\} \right) .
\end{equation}

It is seen that we can choose $B_{1}=0$ and $E_{2}=0$ to have finite limit
when $r\rightarrow 0$. However, for $r\rightarrow \infty $ it can be checked
that $w(r,t)$ diverges and since $g_{0}(r)=1$ this implies that the ratio of 
$\frac{w(r,t)}{g_{0}(r)}$ grows indefinitely. No other choice of constants
suffice to eliminate this divergence. As a result the perturbation converges
for $r\rightarrow 0$ but diverges for $r\rightarrow \infty $ in $d=3$ case.
Let us note that for the time - independent perturbation both, near and far
- region perturbation terms become convergent.

\subsubsection{\protect\bigskip The case for $d=5$ and $d>5$}

Our perturbed line element now for $d=5$ is of the form

\begin{equation}
ds^{2}=f(r,t)\left( dt^{2}-dr^{2}-dx^{2}-dy^{2}\right) -\frac{r^{2}b_{0}^{2}%
}{g(r,t)}d\varphi ^{2}
\end{equation}%
From Eq. (23) and (24) we introduce in analogy, the perturbed functions in
which the unperturbed functions are given by

\begin{eqnarray}
f_{0}(r) &=&2(r^{2}+1) \\
g_{0}(r) &=&4(r^{2}+1)^{2} \\
A_{\varphi }(r) &=&-\frac{B_{0}b_{0}}{8}\frac{1}{(r^{2}+1)}
\end{eqnarray}%
From the Einstein's equations we obtain to the first order in $\epsilon $
the following differential equations for $u(r,t)$ and $w(r,t)$

\begin{gather}
r\frac{\partial ^{2}u\left( r,t\right) }{\partial r^{2}}-\frac{\partial
u(r,t)}{\partial r}-ru(r,t)=0 \\
2(r^{2}+1)\frac{\partial ^{2}u(r,t)}{\partial t^{2}}-\frac{1}{2}\frac{%
\partial ^{2}w(r,t)}{\partial t^{2}}=0
\end{gather}%
Integration of these equations yield

\begin{eqnarray}
u(r,t) &=&e^{-\alpha t}\left[ r(pI_{1}(\alpha r)+qK_{1}(\alpha r))\right] \\
w(r,t) &=&4e^{-\alpha t}\left[ r(pI_{1}(\alpha r)+qK_{1}(\alpha r))\right]
(r^{2}+1)
\end{eqnarray}%
for integration constants $p,q$ and $\alpha $, and $I_{1}(\alpha r)$ and $%
K_{1}(\alpha r)$ \ are the modified Bessel functions of order one. At all
times $t$, to a first order in $\epsilon $ , the locally flat nature of the
metric near $r=0$ remains intact. Also, for all times to a first order in $%
\epsilon $ , the static metric for $r\rightarrow \infty $ will not be
altered. In this case we see that perturbed functions are finite at $r=0,$
and when $r\rightarrow \infty $ they go to zero.

A similar analysis has been carried out for $d=6,7,...$ so that we found the
general solution for the metric functions. In each case we have obtained the
following relations between $w\left( r,t\right) $ and $u\left( r,t\right) $

\begin{gather}
w\left( r,t\right) =\left( d-3\right) \frac{g_{0}\left( r\right) }{%
f_{0}\left( r\right) }u\left( r,t\right) \\
f_{0}\left( r\right) =(d-3)^{\frac{2}{d-3}}(r^{2}+1)^{\frac{2}{d-3}} \\
g_{0}\left( r\right) =(d-3)^{2}(r^{2}+1)^{2}.
\end{gather}%
Solutions for $w\left( r,t\right) $ and $u\left( r,t\right) $ are given in
all dimensions as follow

\begin{eqnarray}
w\left( r,t\right) &=&\left( d-3\right) ^{\frac{3d-11}{d-3}}(r^{2}+1)^{\frac{%
2d-8}{d-3}}u\left( r,t\right) , \\
u\left( r,t\right) &=&e^{-\alpha t}\left( d-3\right) ^{\frac{d-1}{d-3}%
}\left( pI_{1}\left( \alpha r\right) +qK_{1}\left( \alpha r\right) \right) 
\frac{r}{(r^{2}+1)^{\frac{d-5}{d-3}}},
\end{eqnarray}%
in which $p$ and $q$ are constants. These are both finite at $r\rightarrow 0$
and the ratio of them when $r\rightarrow \infty $ go to zero. Now, from
Maxwell's equations we attempt to find the perturbative solution for the
magnetic potential. We take the potential in the form

\begin{equation*}
A_{\varphi }(r,t)=A_{\varphi }(r)+\epsilon a(r,t)
\end{equation*}%
where

\begin{equation}
A_{\varphi }(r)=-\frac{B_{0}b_{0}}{2(d-3)^{2}}\frac{1}{(r^{2}+1)}.
\end{equation}

The differential equation satisfied by $a\left( r,t\right) $ becomes

\begin{equation}
-\frac{\partial ^{2}a\left( r,t\right) }{\partial t^{2}}+\frac{\partial
^{2}a\left( r,t\right) }{\partial r^{2}}+\frac{\left( 3r^{2}-1\right) }{%
r\left( r^{2}+1\right) }\frac{\partial a\left( r,t\right) }{\partial r}+%
\frac{b_{0}B_{0}r}{\left( d-3\right) ^{\frac{d-1}{d-3}}\left( r^{2}+1\right)
^{2}}\frac{d}{dr}\left[ \frac{u\left( r,t\right) }{\left( r^{2}+1\right) ^{%
\frac{2}{d-3}}}\right] =0
\end{equation}%
in which

\begin{equation}
u\left( r,t\right) =e^{-\alpha t}\left( d-3\right) ^{\frac{d-1}{d-3}}\left(
pI_{1}\left( \alpha r\right) +qK_{1}\left( \alpha r\right) \right) \frac{r}{%
(r^{2}+1)^{\frac{d-5}{d-3}}}.
\end{equation}

Upon substitution for $u\left( r,t\right) $ we see that for all dimensions
the equation satisfied by $a\left( r,t\right) $ takes the form

\begin{equation}
-\frac{\partial ^{2}a\left( r,t\right) }{\partial t^{2}}+\frac{\partial
^{2}a\left( r,t\right) }{\partial r^{2}}+\frac{\left( 3r^{2}-1\right) }{%
r\left( r^{2}+1\right) }\frac{\partial a\left( r,t\right) }{\partial r}+%
\frac{b_{0}B_{0}re^{-\alpha t}}{\left( r^{2}+1\right) ^{2}}\frac{d}{dr}\left[
\frac{r}{^{\left( r^{2}+1\right) }}\left( pI_{1}\left( \alpha r\right)
+qK_{1}\left( \alpha r\right) \right) \right] =0.
\end{equation}%
An exact solution for $a\left( r,t\right) $ for all $r$ is not at our
disposal, therefore, we shall search for solutions near $r=0$ and for $%
r\rightarrow \infty $. The solution for the homogenous part is

\begin{equation}
a_{H}\left( r,t\right) =\frac{re^{-\alpha t}}{^{\left( r^{2}+1\right) }}%
\left( C_{1}I_{1}\left( \alpha r\right) +C_{2}K_{1}\left( \alpha r\right)
\right) .
\end{equation}%
Since a particular solution is not available we proceed to study the answers
for a limited case, when $r$ is small $($to order $r)$

\begin{gather}
-\frac{\partial ^{2}a\left( r,t\right) }{\partial t^{2}}+\frac{\partial
^{2}a\left( r,t\right) }{\partial r^{2}}-\frac{1}{r}\frac{\partial a\left(
r,t\right) }{\partial r}=0, \\
a\left( r,t\right) =e^{-\alpha t}r\left( C_{1}I_{1}\left( \alpha r\right)
+C_{2}K_{1}\left( \alpha r\right) \right) .  \notag
\end{gather}%
If we go to higher orders of $r$, $($order $r^{3}$ for instance$)$ we obtain

\begin{equation}
-\frac{\partial ^{2}a\left( r,t\right) }{\partial t^{2}}+\frac{\partial
^{2}a\left( r,t\right) }{\partial r^{2}}+\left[ 4r-\frac{1}{r}\right] \frac{%
\partial a\left( r,t\right) }{\partial r}+b_{0}B_{0}r^{2}\left\{ -p\alpha
\left( -\ln \left( \frac{1}{2}\alpha \right) -\ln r-\gamma \right) +p\alpha -%
\frac{2q}{\alpha }\right\} =0
\end{equation}%
whose solution can be expressed as

\begin{equation}
a\left( r,t\right) =e^{-r^{2}-\alpha t}\left\{ C_{3}Whitta\ker M(-\frac{1}{8}%
\alpha ^{2},\frac{1}{2},2r^{2})+C_{4}Whitta\ker M(-\frac{1}{8}\alpha ^{2},%
\frac{1}{2},2r^{2})\right\} -\frac{B_{0}b_{0}e^{-\alpha t}}{\alpha \left(
-8+\alpha ^{2}\right) ^{2}}\left( \Xi +\Pi \right)
\end{equation}%
in which%
\begin{eqnarray*}
\Xi &=&-\alpha ^{2}r^{2}q\left( -8+\alpha ^{2}\right) \ln \alpha -\alpha
^{2}r^{2}q\left( -8+\alpha ^{2}\right) \ln r+\left[ q\left( \ln 2-\gamma
\right) -p\right] \alpha ^{4}r^{2}, \\
\Pi &=&\left( \left[ \left( 8\gamma -2-8\ln 2\right) r^{2}-2\right]
q+8r^{2}p\right) \alpha ^{2}+16q-16r^{2}q.
\end{eqnarray*}%
Here $p,q,C_{3},C_{4},\alpha $ and $\gamma $ are all constants and $%
WhittakerM\left( -\frac{1}{8}\alpha ^{2},\frac{1}{2},2r^{2}\right) $ stands
for the Whittaker function\cite{9}. For $r\rightarrow 0$ the function $%
a\left( r,t\right) $ is finite. Now, to see the case when $r$ goes to
infinity we solve the differential equation

\begin{equation}
-\frac{\partial ^{2}a\left( r,t\right) }{\partial t^{2}}+\frac{\partial
^{2}a\left( r,t\right) }{\partial r^{2}}=0,\text{ \ \ \ \ (to order }1/r%
\text{)}
\end{equation}%
whose solution reads

\begin{equation}
a\left( r,t\right) =e^{-\alpha t}\left( C_{5}e^{\alpha r}+C_{6}e^{-\alpha
r}\right) .
\end{equation}%
Here $C_{5}$ and $C_{6}$ are new integration constants. It can easily be
checked that the ratio of this solution goes to zero for $r\rightarrow
\infty $ if we choose $C_{5}=0.$

\section{Geodesic motion}

In this section we shall investigate the time-like (for $d\geq 4$) and null
(for $d=3,4$) geodesics by employing our line element given in Eq. (15). For 
$d\geq 4$ we divide the line element by $d\tau $ ($\tau $ is proper time)
and for $d=3$ by $d\lambda $ ($\lambda $ is an affine parameter) so that the
Lagrangian can be expressed in the form

\begin{equation}
L=\left\{ 
\begin{array}{lc}
(d-3)^{\frac{2}{d-3}}(r^{2}+1)^{\frac{2}{d-3}}\left[ \left( \frac{dt}{d\tau }%
\right) ^{2}-\left( \frac{dr}{d\tau }\right) ^{2}-\underset{i=1}{\overset{d-3%
}{\tsum }}\left( \frac{dz_{i}}{d\tau }\right) ^{2}\right] -\frac{%
r^{2}b_{0}^{2}}{(d-3)^{2}(r^{2}+1)^{2}}\left( \frac{d\varphi }{d\tau }%
\right) ^{2}, & \text{ \ \ \ \ \ \ \ \ \ \ \ }d\geq 4 \\ 
e^{cr^{2}}\left[ \left( \frac{dt}{d\lambda }\right) ^{2}-\left( \frac{dr}{%
d\lambda }\right) ^{2}\right] -r^{2}b_{0}^{2}\left( \frac{d\varphi }{%
d\lambda }\right) ^{2}, & \text{ \ \ \ \ \ \ \ \ \ \ \ }d=3%
\end{array}%
\right. .
\end{equation}%
For the equations of motion with constant azimuthal angle ($\varphi =$%
constant for $d\geq 4$) and null geodesics for $d=3$ we obtain%
\begin{equation}
\left\{ 
\begin{array}{ll}
\begin{array}{c}
\frac{dt}{d\tau }=\frac{d_{0}}{(d-3)^{\frac{2}{d-3}}(r^{2}+1)^{\frac{2}{d-3}}%
} \\ 
\frac{dz_{i}}{d\tau }=\frac{d_{i}}{(d-3)^{\frac{2}{d-3}}(r^{2}+1)^{\frac{2}{%
d-3}}}%
\end{array}
& \text{\ \ }d\geq 4 \\ 
\begin{array}{l}
\frac{dt}{d\lambda }=H_{0}e^{-cr^{2}} \\ 
\frac{d\varphi }{d\lambda }=\frac{H_{1}}{r^{2}b_{0}^{2}}%
\end{array}
& d=3%
\end{array}%
\right.
\end{equation}%
in which $d_{0}$ , $d_{i}$, $H_{0}$ and $H_{1}$ are all constants of
integration. From the metric condition we find $\frac{dr}{d\tau }$ (for $%
d\geq 4$, $\varphi =$ constant) and $\frac{dr}{d\lambda }$ (for $d=3$) as
follow

\begin{equation}
\left\{ 
\begin{array}{ll}
\begin{array}{l}
\frac{dr}{d\tau }=\pm \frac{1}{(d-3)^{\frac{2}{d-3}}(r^{2}+1)^{\frac{2}{d-3}}%
}\sqrt{a_{0}^{2}-(d-3)^{\frac{2}{d-3}}(r^{2}+1)^{\frac{2}{d-3}}} \\ 
a_{0}^{2}=d_{0}^{2}-\underset{i=1}{\overset{d-3}{\tsum }}d_{i}^{2}%
\end{array}
& \text{ \ \ \ \ \ \ \ \ \ \ \ \ \ \ \ \ \ \ \ \ \ \ \ \ \ \ \ \ \ \ \ \ \ \
\ }d\geq 4 \\ 
\begin{array}{l}
\frac{dr}{d\lambda }=\pm H_{0}e^{-cr^{2}}\sqrt{1-\frac{H_{2}^{2}}{r^{2}}%
e^{cr^{2}}} \\ 
H_{2}=\frac{H_{1}}{H_{0}b_{0}}%
\end{array}
& \text{ \ \ \ \ \ \ \ \ \ \ \ \ \ \ \ \ \ \ \ \ \ \ \ \ \ \ \ \ \ \ \ \ \ \
\ }d=3%
\end{array}%
\right.
\end{equation}%
In effect, we obtain for $d=4$ the relation between $r$ and $\tau $ but for $%
d=3$ we want to find the relation between $r$ and $\varphi .$

\begin{eqnarray}
&&\left\{ 
\begin{array}{cc}
\begin{array}{c}
\pm \left( \tau -\tau _{0}\right) =(d-3)^{\frac{1}{d-3}}\dint \frac{%
(r^{2}+1)^{\frac{2}{d-3}}dr}{\sqrt{k_{0}^{2}-(r^{2}+1)^{\frac{2}{d-3}}}} \\ 
\left( k_{0}=\frac{a_{0}}{(d-3)^{\frac{1}{d-3}}}\right)%
\end{array}
& \text{\ \ \ \ \ \ \ \ \ \ \ \ \ \ \ \ \ \ \ \ \ \ \ \ \ \ \ \ \ \ \ \ \ \
\ \ \ \ \ \ \ \ \ \ \ \ \ \ \ \ \ \ \ \ \ }d\geq 4%
\end{array}%
\right. \\
&&\text{ \ \ \ \ \ }%
\begin{array}{cc}
\pm \left( \varphi -\varphi _{0}\right) =\frac{1}{H_{0}}\dint \frac{%
e^{cr^{2}}dr}{r\sqrt{r^{2}-H_{2}^{2}e^{cr^{2}}}} & \text{\ \ \ \ \ \ \ \ \ \
\ \ \ \ \ \ \ \ \ \ \ \ \ \ \ \ \ \ \ \ \ \ \ \ \ \ \ \ \ \ \ \ \ \ \ \ \ \
\ \ \ \ \ \ \ \ \ \ \ \ \ \ \ \ \ \ \ }d=3%
\end{array}%
\end{eqnarray}%
in which, $\tau _{0}$ and $\varphi _{0}$ are initial constants and we impose
the restrictions so that

\begin{equation}
\left\{ 
\begin{array}{cc}
(r^{2}+1)^{\frac{2}{d-3}}<k_{0}^{2} & \text{ \ \ \ \ \ \ \ \ \ \ \ \ \ \ \ \
\ \ \ \ \ \ \ \ \ \ \ \ \ \ \ \ \ \ \ \ \ \ \ \ \ \ \ \ \ \ \ \ \ \ \ \ \ \
\ \ \ \ \ \ \ \ \ \ \ \ \ \ \ \ \ \ \ \ \ \ \ \ \ \ \ \ \ \ \ \ \ \ \ \ \ \
\ \ \ }d\geq 4 \\ 
\frac{H_{2}^{2}}{r^{2}}e^{cr^{2}}<1 & \text{ \ \ \ \ \ \ \ \ \ \ \ \ \ \ \ \
\ \ \ \ \ \ \ \ \ \ \ \ \ \ \ \ \ \ \ \ \ \ \ \ \ \ \ \ \ \ \ \ \ \ \ \ \ \
\ \ \ \ \ \ \ \ \ \ \ \ \ \ \ \ \ \ \ \ \ \ \ \ \ \ \ \ \ \ \ \ \ \ \ \ \ \
\ \ }d=3%
\end{array}%
\right. .
\end{equation}%
For $d=4$ and $5$, we have exact integrals given by

\begin{equation}
\tau -\tau _{0}=\left\{ 
\begin{array}{ll}
-\frac{1}{3}r\sqrt{k_{0}^{2}-(r^{2}+1)^{2}}+\frac{\sqrt{1+\frac{r^{2}}{%
\left( k_{0}+1\right) }}\left( k_{0}-1\right) \sqrt{1-\frac{r^{2}}{\left(
k_{0}-1\right) }}}{3\sqrt{k_{0}^{2}-(r^{2}+1)^{2}\sqrt{k_{0}-1}}}\Omega & 
\text{\ \ }d=4 \\ 
\sqrt{2}\left[ -\frac{1}{2}r\sqrt{k_{0}^{2}-(r^{2}+1)}+\frac{1}{2}\left(
k_{0}^{2}+1\right) \arctan \left( \frac{r}{\sqrt{k_{0}^{2}-(r^{2}+1)}}%
\right) \right] & \text{ \ }d=5%
\end{array}%
\right.
\end{equation}%
in which 
\begin{equation*}
\Omega =\left( k_{0}^{2}-2k_{0}\right) \text{EllipticF}\left( \frac{r}{\sqrt{%
k_{0}-1}},\sqrt{-1+\frac{2}{\left( k_{0}+1\right) }}\right) +\left(
2k_{0}+2\right) \text{EllipticF}\left( \frac{r}{\sqrt{k_{0}-1}},\sqrt{-1+%
\frac{2}{\left( k_{0}+1\right) }}\right) .
\end{equation*}%
Fig.s 1a and 1b depict the behaviors of (63) (for $d=4$) and (66) (for $d=5$%
) , respectively. Now, we wish to consider the $d=4$ null geodesics as well.
The line element is

\begin{equation}
ds^{2}=(r^{2}+1)^{2}\left( dt^{2}-dr^{2}-dz^{2}\right) -\frac{r^{2}b_{0}^{2}%
}{(r^{2}+1)^{2}}d\varphi ^{2}
\end{equation}%
Eq.s of motion imply for the affine parameter $\lambda $

\begin{equation}
\frac{dt}{d\lambda }=\frac{\alpha _{0}}{(r^{2}+1)^{2}},\text{ \ \ \ }\frac{dz%
}{d\lambda }=\frac{\beta _{0}}{(r^{2}+1)^{2}},\text{ \ \ \ }\frac{d\varphi }{%
d\lambda }=\gamma _{0}\frac{(r^{2}+1)^{2}}{r^{2}}
\end{equation}%
in which $\alpha _{0}$, $\beta _{0}$ and $\gamma _{0}$ are integration
constants. We note that (67) and (68) correspond to Eq.s (1) and (6) of \cite%
{3}, respectively. From the null-metric condition $ds^{2}=0$ and upon
shifting the independent variable to $\varphi $\ we obtain

\begin{equation}
\pm \left( \varphi -\varphi _{0}\right) =\dint \frac{(r^{2}+1)^{4}dr}{r\sqrt{%
\lambda _{0}^{2}r^{2}-b_{0}^{2}(r^{2}+1)^{4}}},
\end{equation}%
where we have introduced the constant $\lambda _{0}$ as $\lambda _{0}^{2}=%
\frac{\alpha _{0}^{2}-\beta _{0}^{2}}{\gamma _{0}^{2}}$. Fig.s 2a and 2b are
the plots corresponding to (64) and (69), respectively.

Finally, we study the $4$-dimensional time-like geodesics of a charged
particle with charge $e$ whose Lagrangian is given by

\bigskip

\begin{equation}
L=(r^{2}+1)^{2}\left[ \left( \frac{dt}{d\tau }\right) ^{2}-\left( \frac{dr}{%
d\tau }\right) ^{2}-\left( \frac{dz}{d\tau }\right) ^{2}\right] -\frac{%
r^{2}b_{0}^{2}}{(r^{2}+1)^{2}}\left( \frac{d\varphi }{d\tau }\right) ^{2}-%
\frac{eB_{0}b_{0}}{2}\frac{1}{(r^{2}+1)}\left( \frac{d\varphi }{d\tau }%
\right) .
\end{equation}%
The Euler-Lagrange equations of motion yield

\begin{equation}
\frac{dt}{d\tau }=\frac{l_{0}}{(r^{2}+1)^{2}},\text{ }\frac{dz}{d\tau }=%
\frac{\mu _{0}}{(r^{2}+1)^{2}},\text{ }\frac{d\varphi }{d\tau }=\frac{1}{2}%
\frac{\left( -\sigma _{0}r^{2}-\sigma _{0}+b_{0}a_{1}\right) (r^{2}+1)}{%
b_{0}^{2}r^{2}},\text{ }a_{1}=-\frac{eB_{0}}{2},
\end{equation}%
for the integration constants $l_{0}$, $\mu _{0}$ and $\sigma _{0}$. Since
we shall be interested only in the $r\left( \tau \right) $ behavior of the
motion we derive the second order equation as follows

\begin{gather}
-2(r^{2}+1)^{2}\frac{d^{2}r}{d\tau ^{2}}+\left[ -8r(r^{2}+1)+\frac{4r}{%
(r^{2}+1)^{3}}\left( r^{8}+4r^{6}+6r^{4}+4r^{2}+1\right) \right] \left( 
\frac{dr}{d\tau }\right) ^{2} \\
+\frac{1}{2r^{3}b_{0}^{2}(r^{2}+1)^{3}}\left[ \sigma _{0}^{2}r^{10}+3\sigma
_{0}^{2}r^{8}+r^{6}\left( 2\sigma _{0}^{2}+b_{0}^{2}a_{1}^{2}\right)
-r^{4}\left( 2\sigma _{0}^{2}+b_{0}^{2}\left[ -3a_{1}^{2}+2\mu
_{0}^{2}-2l_{0}^{2}\right] \right) \right.  \notag \\
\left. -r^{2}\left( 3\sigma _{0}^{2}-3a_{1}^{2}b_{0}^{2}\right) -\sigma
_{0}^{2}+a_{1}^{2}b_{0}^{2}\right] =0.  \notag
\end{gather}%
For a set of chosen constants and initial values we plot the behavior of $%
r\left( \tau \right) $ as depicted in Fig. 3. Our overall analysis shows
that irrespective of the initial conditions $r\left( \tau \right)
\rightarrow \infty $ , with the increasing proper time.

\section{\textbf{CONCLUSION}}

We rederive the family of cylindrically symmetric magnetic universes in a
particular metric ansatz which is conformally flat on each constant
azimuthal angle. These are non-black hole solutions where unlike their
spherical counterparts the gravity of magnetic fields is not strong enough
to make black holes. The energy conditions (in the Appendix) of the magnetic
field are satisfied only in particular dimensions. Being inspired by the
stability properties of the original $4$-dimensional Melvin's magnetic
universe and those of Gibbons and Wiltshire we prove also the stability of
present universes in a different gauge and in all dimensions including $d=3$%
. Small radial perturbations of metric functions and the magnetic field
(which automatically yields an electric field in accordance with the Maxwell
equations) result in convergent expansions. Stability of the $3$-dimensional
case which was not considered in previous studies turns out to be weaker
(i.e. convergence of perturbations are not satisfied simultaneously at $r=0$%
\ and at $r\rightarrow \infty $).

Geodesics show numerically that in running proper time uncharged particles
are confined while null geodesics spiral around the center. Exact,
particular geodesics in terms of the elementary functions are available in $%
d=5,$ whereas in $d=4$ we have elliptic functions.

\textbf{Acknowledgment:} We wish to thank S. Habib Mazharimousavi for much
valuable discussions. \ \ \ \ \ \ \ \ \ \ \ \ \ \ \ \ \ \ \ \

\textbf{Apendix A}

\textbf{Energy conditions}

When a matter field couples to any system, energy conditions must be
satisfied for physically acceptable solutions. We follow the steps as given
in\cite{10} .

\textit{Weak Energy Condition (WEC):}

\bigskip The WEC states that

\begin{eqnarray}
\rho &\geq &0  \TCItag{A1} \\
\rho +p_{i} &\geq &0  \notag
\end{eqnarray}

In which $\rho $ is the energy density and \ $p_{i}$are the principle
pressure given by

\begin{eqnarray}
\rho &=&T_{0}^{0}  \TCItag{A2} \\
p_{i} &=&-T_{i}^{i},i=1,2,\cdots ,(d-1)  \notag
\end{eqnarray}

The WEC conditions are trivially satisfied.

\textit{Strong Energy Condition (SEC):}

This condition states that

\begin{eqnarray}
\rho +\underset{i=1}{\overset{d-1}{\tsum }}p_{i} &\geq &0  \TCItag{A3} \\
\rho +p_{i} &\geq &0  \notag
\end{eqnarray}

For d=3, it means that

\begin{eqnarray}
\rho +\underset{i=1}{\overset{d-1}{\tsum }}p_{i} &\geq &0\Rightarrow
3T_{0}^{0}\geq 0  \TCItag{A4} \\
\rho +p_{i} &\geq &0\Rightarrow 2T_{0}^{0}\geq 0  \notag
\end{eqnarray}%
which are satisfied. For $4\leq d\leq 6$\ it is also satisfied because $\rho
+\underset{i=1}{\overset{d-1}{\tsum }}p_{i}=-(d-6)T_{0}^{0}$. It can easily
be seen also that for $7\leq d$

\begin{eqnarray}
\rho +\underset{i=1}{\overset{d-1}{\tsum }}p_{i} &\geq &0\Rightarrow
-T_{0}^{0}\geq 0  \TCItag{A5} \\
\rho +p_{i} &\geq &0\Rightarrow T_{0}^{0}\geq 0  \notag
\end{eqnarray}%
i.e. the SEC is violated.

\bigskip

\textit{Dominant Energy Condition (DEC):}

In accordance with DEC, the effective pressure should not be negative. This
amounts to

\bigskip 
\begin{equation}
p_{eff}=\frac{1}{d-1}\underset{i=1}{\overset{d-1}{\tsum }}T_{i}^{i}=\frac{%
(d-5)}{d-1}T_{0}^{0}\geq 0  \tag{A6}
\end{equation}

For having $p_{eff}\geq o$ \ it is clear that $d\geq 5.$

\bigskip \textit{Causality Condition:}

In addition to the energy conditions one can impose the causality condition

\begin{equation}
0\leq \frac{p_{eff}}{\rho }<1  \tag{A7}
\end{equation}

This implies that

\begin{equation}
0\leq \frac{(d-5)}{d-1}<1  \tag{A8}
\end{equation}%
which is satisfied for $d\geq 5.$

\bigskip

\textbf{Figure captions }

Fig. 1 : Radial distance behavior as a function of proper time for
time-like, neutral particle geodesics. From Eq. (63) in the text, the
behavior of $r\left( \tau \right) $ as the proper time $\tau $ runs from
zero to infinity. The starting points are chosen such that $r=0$ at $\tau =0$%
, in both $d=4$ (Fig. 1a) and $d=5$ (Fig. 1b)cases. The fact that $r$ is
confined is clearly seen from these plots. This particular property is
already implied from Eq. (65). Let us note that in both cases for simplicity
we choose $\left( +\right) $ sign and $\tau _{0}=0$. Further, we choose the
constant $k_{0}=2.$

Fig. 2 :\ Radial behavior as a function of the azimuthal angle $\varphi $
for null geodesics in $d=3$ (Fig 2a) and $d=4$ dimensional (Fig. 2b)
magnetic universes. The horizontal axis $x$ $\left( =r\cos \varphi \right) $
and vertical axis $y$ $\left( =r\sin \varphi \right) $ are plotted
numerically in each case from the expressions given in Eq. (64) and Eq. (69)
. For simplicity we choose the constants $H_{2}=0.01$, $b_{0}=0.01$ , $%
H_{0}=1$, $c=1$ and $\lambda _{0}=1.$ Let us note that we have chosen the $%
\left( +\right) $ sign in both (64) and (69) , which give outward orbits
around the center. Obviously, the choice $\left( -\right) $ should yield
inward orbits.

Fig. 3 :\ This is a numerical plot of the intricate second order
differential equation (72) in $d=4$. For technical reason we choose the
constants $a_{1},$ \ $b_{0},$ $l_{0},$ $\mu _{0}$ and $\sigma _{0}$ all
equal to one. Two different initial conditions are displayed (A and B) which
reveal the pattern of increasing radial distance in proper time.

\bigskip

\bigskip

\bigskip


\begin{thebibliography}{99}
\bibitem{1} M. A. Melvin, Phys. Lett. 8, 65 (1964);

M. A. Melvin, Phys. Rev. 139, B225 (1965).

\bibitem{2} D. Garfinkle and M. A. Melvin, Phys. Rev. D 50, 3859 (1994).

\bibitem{3} K. S. Thorne, Phys. Rev.139, B 244 (1965);

K. S. Thorne, Phys.Rev.138, B251 (1965).

\bibitem{4} Mauricio Cataldo and Patricio Salgado, Phys. Rev. D 54, 2971
(1996);

E. W. Hirschmann and D. L. Welch, Phys .Rev. D 53, 5579 (1996).

\bibitem{5} Ajanta Das and A. Banerjee, Astrophysics and Space Science 268,
425 (1999);

T. Dereli, A. Eris, A. Karasu, Nuovo.Cimento B 93, 102 (1989).

\bibitem{6} S. S. Xulu, Int .J. Mod. Phys. A15 4849 (2000).

\bibitem{7} N. Okuyama and K. Maeda, Phys. Rev. D 67, 104012 (2003);

S. H Mazharimousavi, and M. Halilsoy, Phys. Lett. B 659, 471 (2008);

S. H Mazharimousavi, and M. Halilsoy, Phys. Lett. B 665, 125 (2008);

G. W. Gibbons and C. A. R. Herdeiro, Class. Quant. Grav.18,1677 (2001);

A. Tseytlin, Phys. Lett. B 346, 55 (1995);

F. Dowker, J.Gauntlett, S. Giddings, G. Horowitz, Phys. Rev. D 50, 2662
(1994);

F. Dowker, J.Gauntlett, S. Giddings, G. Horowitz, Phys. Rev. D 52, 6929
(1995).

\bibitem{8} G. W. Gibbons and D. L. Wiltshire, Nucl. Phys. B287, 717 (1987);
(We thank Professor Gibbons for informing us about this reference.)

\bibitem{9} G. B. Arfken and H. J. Weber, Mathematical Methods for Physics,
Fifth edition, Chapter 13, page 858, printed in the USA (2001).

\bibitem{10} S. W. Hawking and G. F. R. Ellis, The Large Scale Structure of
Space-Time, Cambridge University Press (1973);

M. Salgado, Class. Quant. Grav. 20, 4551, (2003);

S. H Mazharimousavi, O. Gurtug and M. Halilsoy, Int. J. Mod. Phys. D 18,
2061 (2009).
\end{thebibliography}
\end{document}